\documentclass[sigchi]{acmart}

\AtBeginDocument{%
  \providecommand\BibTeX{{%
    \normalfont B\kern-0.5em{\scshape i\kern-0.25em b}\kern-0.8em\TeX}}}

\setcopyright{acmcopyright}

\copyrightyear{2023}
\acmYear{2023}
\setcopyright{acmlicensed}\acmConference[CHI '23]{Proceedings of the 2023 CHI Conference on Human Factors in Computing Systems}{April 23--28, 2023}{Hamburg, Germany}
\acmBooktitle{Proceedings of the 2023 CHI Conference on Human Factors in Computing Systems (CHI '23), April 23--28, 2023, Hamburg, Germany}
\acmPrice{15.00}
\acmDOI{10.1145/3544548.3581271}
\acmISBN{978-1-4503-9421-5/23/04}

\usepackage{multirow}
\usepackage{cancel}
\usepackage{soul}
\usepackage{balance}

\definecolor{NewContentColor}{HTML}{C7372F}  

\definecolor{DeleteColor}{HTML}{6495ED}  
    {} 

\newcommand{\tm}[1]{}
\newcommand{\cb}[1]{}
\newcommand{\sk}[1]{}
\newcommand{\delete}[1]{}
\newcommand{\new}[1]{{#1}}

\newcommand{\activity}[1]{#1}    

\definecolor{DSBlue}{HTML}{87CEFA}          
\definecolor{DJGreen}{HTML}{7FC97F}         
\definecolor{RelatedPurple}{HTML}{EF7DBE}   

\newcommand{\activityDS}[1]{\colorbox{DSBlue}{#1}}    
\newcommand{\activityDJ}[1]{\colorbox{DJGreen}{#1}}    
\newcommand{\discrepancy}[1]{\textsc{#1}} 

\newcommand{\SmithHopkins}[0]{P02}      
\newcommand{\McDonald}[0]{P03}          
\newcommand{\Cano}[0]{P05}              
\newcommand{\Rocha}[0]{P06}             
\newcommand{\Arthur}[0]{P07}            
\newcommand{\Weisz}[0]{P08}             
\newcommand{\McKie}[0]{P09}             
\newcommand{\Bradshaw}[0]{P12}          
\newcommand{\Pantazi}[0]{P13}           
\newcommand{\Jones}[0]{P14}             
\newcommand{\Akin}[0]{P18}              
\newcommand{\Kanik}[0]{P20}             
\newcommand{\Zhang}[0]{P30}             
\newcommand{\Jingnan}[0]{P36}           

\newcommand{\DSPapersTotal}[0]{31}     
\newcommand{\DSPaperCount}[0]{16}      
\newcommand{\DSExcerpts}[0]{150}       

\newcommand{\DSActivityCount}[0]{30}   
\newcommand{\DSIssueCount}[0]{11}     
\newcommand{\DSCodes}[0]{41}           

\newcommand{\N}[0]{36}                         

\newcommand{\InterviewActivityCount}[0]{13}    
\newcommand{\InterviewIssueCount}[0]{15}       
\newcommand{\InterviewCodes}[0]{28}            

\newcommand{\InitialActivityCount}[0]{43}      
\newcommand{\InitialIssueCount}[0]{26}         
\newcommand{\ActivityCount}[0]{23}             

\newcommand{\DQTaxonomyCount}[0]{16}   

\newcommand{\DQIssueCountRaw}[0]{330}      
\newcommand{\DQIssueCountExcluded}[0]{15}  
\newcommand{\DQIssueCount}[0]{315}         
\newcommand{\ClusterCount}[0]{45}          
\newcommand{\FinalIssueCount}[0]{60}       

\newcommand{\DQIssueCountUnique}[0]{16}             
\newcommand{\InterviewIssueCountUnique}[0]{13}      
                                                    
\newcommand{\IssueCountOverlap}[0]{31}             

\newcommand{\HorrorStories}[0]{69}             
\newcommand{\HorrorChallenges}[0]{63}          
\newcommand{\HorrorSnippets}[0]{104}           
\newcommand{\InterviewSnippets}[0]{566}        

\newcommand{\SuppActivities}[0]{\href{https://osf.io/vfzpg}{Supp.~Section/Sheet~3}}
\newcommand{\SuppIssues}[0]{\href{https://osf.io/x8rkv}{Supp.~Section/Sheet~4}}
\newcommand{\SuppTools}[0]{\href{https://osf.io/j5fgb}{Supp.~Section/Sheet~5}}
\newcommand{\SuppParticipants}[0]{\href{https://osf.io/q8kun}{Supp.~Sheet~2}}
\newcommand{\SuppDQTaxonomyCounts}[0]{\href{https://osf.io/t2dya}{Supp.~Section/Sheet~1}}
\newcommand{\SuppDSPapers}[0]{\href{https://osf.io/t2dya}{Supp.~Section/Sheet~1}}
\newcommand{\SuppDataIssues}[0]{\href{https://osf.io/keq86}{Supp.~Section~4}}
\newcommand{\SuppInterviewScript}[0]{\href{https://osf.io/n4q8b}{Supp.~Section~2}}
\newcommand{\SuppIntegrationChallenges}[0]{\href{https://osf.io/tfdzu}{Supp.~Sheet~6}}
\newcommand{\SuppAbandon}[0]{\href{https://osf.io/d8s57}{Supp.~Section~6}}

\makeatletter
\def\testclr#1#{\@testclr{#1}}
\def\@testclr#1#2{{\fboxsep\z@\fbox{\colorbox#1{#2}{\phantom{XX}}}}}

\makeatother

\begin{document}

\title[Dirty Data in the Newsroom]{Dirty Data in the Newsroom: Comparing Data Preparation in Journalism and Data Science}

\author{Stephen Kasica}
\email{kasica@alumni.ubc.ca}
\orcid{0000-0002-9775-9720}
\affiliation{%
  \institution{The University of British Columbia}
  \streetaddress{201-2366 Main Mall}
  \city{Vancouver}
  \state{BC}
  \postcode{V6T 1Z4}
  \country{Canada}}
\author{Charles Berret}
\email{charles.berret@liu.se}
\orcid{0000-0002-2796-6820}
\affiliation{%
  \institution{Link{\"o}ping University}
  \streetaddress{Kopparhammaren 2, MIT/ITN}
  \postcode{601 74}
  \city{Norrk{\"o}ping}
  \country{Sweden}}
\author{Tamara Munzner}
\email{tmm@cs.ubc.ca}
\orcid{0000-0002-3294-3869}
\affiliation{%
  \institution{The University of British Columbia}
  \streetaddress{201-2366 Main Mall}
  \city{Vancouver}
  \state{BC}
  \postcode{V6T 1Z4}
  \country{Canada}}

\begin{abstract}
The work involved in gathering, wrangling, cleaning, and otherwise preparing data for analysis is often the most time consuming and tedious aspect of data work. Although many studies describe data preparation within the context of data science workflows, there has been little research on data preparation in data journalism. We address this gap with a hybrid form of thematic analysis that combines deductive codes derived from existing accounts of data science workflows and inductive codes arising from an interview study with 36 professional data journalists. We extend a previous model of data science work to incorporate detailed activities of data preparation. We synthesize 60 dirty data issues from 16 taxonomies on dirty data and our interview data, and we provide a novel taxonomy to characterize these dirty data issues as discrepancies between mental models. We also identify four challenges faced by journalists: diachronic, regional, fragmented, and disparate data sources.
\end{abstract}




\maketitle  

\section{Introduction}

A large body of research addresses data preparation in data science, where studies show the work of wrangling, cleaning, and otherwise preparing data for analysis can be responsible for 80\% of the time and cost of data warehousing projects~\cite{dasu_exploratory_2003}. We seek to understand how closely the abundant research on data scientists applies to data journalists, who use computational tools and techniques to leverage data for the production of news. Data journalists often gather and analyze datasets using structured, quantitative information as an additional source to fact check claims, supplement material gathered through traditional methods like interviewing, and support in-depth investigations. Previous studies have suggested a close relationship exists between data journalists and data scientists with regards to tool usage, data sources, and work practices~\cite{kasica_table_2021, showkat_where_2021}.

Although we see clear parallels between the activities of these two groups, the relationship between data journalism and the adjacent field of data science requires further study. 
While anecdotal evidence points to the prevalence and difficulty of data preparation in data journalism~\cite{hutchins_data_2020}, we lack empirical data on the specific challenges faced by data journalists in comparison to data scientists. A significant body of interview-based research has attempted to understand the daily workflows of data scientists by studying the lived experience of practitioners across diverse domains~\cite{kim_emerging_2016,kim_data_2018,kaggle_state_2019,mao_how_2019,wang_human-ai_2019,wang_how_2019,muller_how_2019,zhang_how_2020,kandogan_data_2014,kandel_enterprise_2012,rule_exploration_2018,alspaugh_futzing_2019,milani_visualization_2020,wongsuphasawat_goals_2019}. \delete{However, none of these include any data journalists as participants.}\new{However, none of these studies include a journalist among their participants.} In this paper, we will use \delete{\textbf{data worker}}\emph{data worker} as an umbrella term to mean both data scientists and data journalists, particularly when emphasizing their commonalities. Considering the needs of data workers, broadly construed, may help to narrow \delete{``the research to reporting gap``}\new{the research to reporting gap}~\cite{stray_making_nlp_2017}, the \new{frequent} ineffectiveness of tools and techniques proposed by computer scientists when applied to the problems that journalists actually encounter.

We examine the extent to which accounts of data preparation among data scientists match the preparatory process of data journalists, featuring a semi-structured interview study with \N~data journalists analyzed qualitatively. Our hybrid thematic analysis incorporates both deductive \delete{\textnormal{a priori}}\new{\emph{a priori}} codes and inductive \delete{\textnormal{a posteriori}}\new{\emph{a posteriori}} codes~\cite{swain_hybrid_2018}. We construct an initial \delete{a priori}\new{\emph{a priori}} codeset by analyzing \DSPaperCount~research papers on data science workflows that address data preparation, allowing us to note where our  \delete{a posteriori}\new{\emph{a posteriori}}  interview findings diverge or overlap with previous work. We also analyze \DQTaxonomyCount~taxonomies of dirty data from the database and data warehousing literature to compare and contrast the conventional wisdom in those fields with our interview findings. Our work provides four contributions. First, the results of a semi-structured interview study with 36 data journalists: This interview study addresses a longstanding research gap, offering a novel perspective of data journalism that contributes to a more complete and pluralistic understanding of data work as a whole~\cite{dignazio_data_2020}. The results of this study are two-fold: a set of \delete{\textbf{activities}}\new{activities} undertaken by data journalists during data preparation, and the set of data quality \delete{\textbf{issues}}\new{issues} they face. We situate these results within the research literature through additional analysis, leading to three additional formalism contributions:

Second, an augmented model of preparatory activities: In Section~\ref{sec:activities}, we present a 
synthesis set of \ActivityCount~data preparation \delete{\emph{activities}}\new{activities}. We combine and consolidate the activities revealed in our interview study with those articulated in previous work, which we identified through a thematic analysis of \DSPaperCount~papers documenting data science workflows. We situate these activities by extending the process model of data science work proposed by Crisan et al.~\cite{crisan_passing_2020} to an additional level of fine-grained activities, grouping them according to preparation \delete{sub-process}\new{subprocess}es (initiate, gather, create, profile, wrangle) and communication \delete{sub-process}\new{subprocess}es (disseminate and document).

Third, a new model-discrepancy taxonomy of dirty data issues: In Section~\ref{sec:model-discrepancy-framework}, we propose a new taxonomy to classify \FinalIssueCount~dirty data issues as discrepancies between mental models among different \delete{data journalists and data scientists}\new{data workers}. Our \new{dirty-data} taxonomy features a two-dimensional design space, with an axis of four data objects (table, item, attribute, and value) and an axis of six data qualities (accuracy, completeness, form, granularity, relation, and semantics). It reconciles the top-down, domain-focused perspective of our practitioner participants and the bottom-up, theory-focused perspective commonly used by computer science researchers to define and describe data issues. We incorporate the latter by analyzing \DQIssueCount~instances of dirty data documented in \DQTaxonomyCount~previous data warehousing papers. 

Finally, four challenges in multi-table data integration: In Section~\ref{sec:integration}, we identify four data integration challenges: \emph{regional} inconsistencies from independent spatially dispersed data sources, \emph{diachronic} inconsistencies from tables recording the same phenomena that evolve over time, \emph{fragmented} tables containing different yet related items that must be re-assembled, and \emph{disparate} tables that are topically dissimilar yet must be related. We identify these challenges from our analysis of the multi-table data integration \delete{``nightmare stories''}\new{nightmare stories} described by participants, which illuminate both activities and issues.

We also provide extensive supplemental materials documenting our qualitative process, with both backing spreadsheets and detailed prose discussions about each table at \new{\url{https://osf.io/nbtvm}}.

\section{Related Work}

Our interdisciplinary work is broadly related to studies in journalism and mass communication as well as human-computer interaction. We divide the most relevant areas of research into data preparation in data science, measuring and classifying errors in data, and data preparation in journalism.

\subsection{Data preparation in data science}

The importance and ubiquity of data preparation is well known in data science~\cite{aho_demystifying_2020}. Many researchers proposing end-to-end process models for conducting data science include specific stages for data preparation~\cite{wirth_crisp-dm_2000, crisan_passing_2020} or synonymous labels such as such as wrangling~\cite{kandel_enterprise_2012, wongsuphasawat_goals_2019, severtson_what_2017}, scrubbing~\cite{mason_taxonomy_2010}, or preprocessing~\cite{feyyad_data_1996}. While this body of research characterizes the entire data science process, our work focuses exclusively on data preparation. We choose to build on the model of data science work from Crisan et al.~\cite{crisan_passing_2020} for three reasons. First, this model provides a clear distinction between preparation and analysis \delete{processes}. Second, its comprehensiveness exceeds other models due to being synthesized from a systematic literature review. Finally, this model provides a sufficiently high-level characterization of workflows that it generalizes to our broad category of data workers, \new{which} \delete{including}\new{includes} data journalists. Our study provides additional levels of detail for two higher-order processes identified in their model, preparation and communication. We do not address its other stages, namely analysis and deployment.

Many researchers report tasks, challenges, pain points, and tool usage during data preparation via broader inquiries into the general workflows of data \new{scientists}. Artificial intelligence practitioners working in high-stakes domains discard potentially valuable data due to missing metadata~\cite{sambasivan_everyone_2021}. Muller et al.~\cite{muller_how_2019} characterizes data wrangling along dimensions of intervention. While studying the workflows of data scientists in software engineering teams, Kim et al.~\cite{kim_emerging_2016, kim_data_2018} identifies specific participant activities typically associated with data preparation, such as merging, cleaning, and shaping data. We identify a larger set of preparation activities and discuss those activities within the context of data issues.

Integrating data is a common challenge during data preparation~\cite{kandel_enterprise_2012, crisan_fits_2021, kandogan_data_2014, muller_how_2019, kim_data_2018}. Kandel et al.~\cite{kandel_enterprise_2012} finds that missing and inconsistent identifiers between tables impede data integration. Kandogan et al.~\cite{kandogan_data_2014} addresses the necessity and absence of semantic metadata when integrating tables. Our study corroborates these findings in the context of data journalism and identifies further data issues that make this activity challenging.

Several \delete{other} studies examine the use of visualization tools \new{for data preparation} and the role of exploration in data science workflows \delete{for data preparation}. Wongsuphasawat et al.~\cite{wongsuphasawat_goals_2019} finds that assessing the quality of data is an exploratory goal. Batch \& Elmqvist~\cite{batch_interactive_2018} identifies a ``visualization gap'', meaning that visualization is under-utilized beyond the final checking and dissemination stages despite research showing its benefits. Milani et al.~\cite{milani_visualization_2020} observes this visualization gap among data analysts cleaning and standardizing data. Alspaugh et al.~\cite{alspaugh_futzing_2019} finds that exploratory activity in the overall data analysis process involves understanding semantics, identifying structure, characterizing data, and assessing quality during data preparation. Our study also finds a visualization gap when assessing the quality of raw data, and identifies other areas where visualization could be leveraged in data work.

\subsection{Dirty data} \label{sec:related-dirty-data}

The term \delete{\textbf{dirty data}}\new{``dirty data''} is used at two levels. At a low level, it means problematic individual items within a dataset, and at a higher level it means the properties of a dataset that degrade its quality; we use the latter definition. While dirty data is an under-researched subject in journalism and mass communication~\cite{messner_journalisms_2017}, database researchers have studied this subject in depth. Companies can lose 20\% of revenue from errors that propagate through a system due to dirty data~\cite{feldman_data_2020}. There are many descriptive models for dirty data to frame the data issues that a proposed technical contribution addresses, evaluate data cleaning tools, or measure data warehouse quality~\cite{chatterjee_data_1991, kim_classifying_1991, rahm_data_2000, kim_taxonomy_2003, muller_problems_2003, barateiro_survey_2005, oliveira_formal_2005, oliveira_taxonomy_2005, li_rule_2011, gschwandtner_taxonomy_2012, de_almeida_taxonomy_2013, roeder_towards_2020}.

Different taxonomies frame the same atomic types of dirty data according to different schemes. One common scheme involves structural vs.~semantic distinctions between dirty data. Chatterjee \& Segev~\cite{chatterjee_data_1991} applies this scheme to catalog problems arising from data heterogeneity,\delete{meaning} the differences between independently maintained data stores. Likewise, Kim \& Seo~\cite{kim_classifying_1991} uses this scheme in a taxonomy of \new{multi-database system} conflicts. \delete{in multi-database systems} Finally, Barateiro \& Galhardas~\cite{barateiro_survey_2005} uses this convention when organizing data quality issues by which one to evaluate using various data cleaning tools.

Another scheme involves high-level distinctions between issues involving a single data source or multiple data sources. Gschwandtner et al.~\cite{gschwandtner_taxonomy_2012} applies this scheme to classify types of dirty time-oriented data. Oliveria et al.~\cite{oliveira_taxonomy_2005} derives a taxonomy from an analysis of production databases in the retail sector, distinguishing between single vs. multiple source issues. Rahm \& Do~\cite{rahm_data_2000} incorporates both schemes: classifying data quality problems according to single source vs. multiple source origins and at the schema vs. instance (semantic) level. Our framework also addresses issues involving multiple tables but does not make high-level distinctions between single-source and multiple-source problems. We find that many issues involving one data source are compounded when working with multiple data sources.

Kim et al.~\cite{kim_taxonomy_2003} contributes an extensive taxonomy of dirty data organized into eight permutations of three binary categories: missing, wrong, or unusable data. From 33 types of dirty data, at least 25 require some form of human intervention. Our work further describes the ways in which data workers intervene to remedy these and other issues.

Outside of data warehousing research, taxonomies of dirty data generally provide a high-level classification of issue categories. Dasu \& Johnson~\cite{dasu_exploratory_2003} names four challenges in exploratory data mining: heterogeneity, quality, scale, and paradigm. Hellerstein~\cite{hellerstein_quantitative_2008} reports four sources of data errors in a survey of quantitative data cleaning strategies: entry, measurement, distillation, and integration. When detecting anomalies in univariate data, Kandel et al.~\cite{kandel_profiler_2012} identifies five specific categories of data anomalies guided by these taxonomies: missing, erroneous, inconsistent, extreme, and key violations. Finally, Wickham~\cite{wickham_tidy_2014} describes the five common problems with messy data involving mismatches between data variables and observations with their representation in rows, columns, and tables that are addressable through data tidying procedures.

Our work is most similar to a group of past taxonomies that enumerate properties of data quality and frame dirty data as a threat to these properties. The taxonomy from de Almeida et al.~\cite{de_almeida_taxonomy_2013} organizes data quality problems into five categories of compromised data and maps each problem to where it manifests in a model of multidimensional data warehousing. Likewise, Oliveira, et al. ~\cite{oliveira_formal_2005} provides formal definitions of data quality issues according to the multidimensional model. M{\"u}ller \& Freytag~\cite{muller_problems_2003} classifies data anomalies into one of three categories that affect nine data quality criteria. Finally, Li et al.~\cite{li_rule_2011} proposes a rule-based taxonomy that classifies dirty data into violations of 13 data quality rules. In contrast, our taxonomy classifies dirty data according to six data qualities and a simplified \delete{four-object}\new{four object} model that more accurately describes the way data journalists discuss issues with data.

\subsection{Data preparation in journalism}

Preparing data has been an important part of data-oriented newswork long before the term ``data journalism'' was coined in the early 2000s~\cite{parasie_computing_2022}. Understanding the context around data, or lack thereof, is a longstanding and important part of data preparation in economic journalism~\cite{arrese_beginning_2022}. Professional organizations for data-oriented newswork have been formalizing and disseminating practical knowledge on data cleaning since at least the early 1990s~\cite{parasie_computing_2022}. Today, data wrangling is one topic where applied artificial intelligence research can have an immediate impact for journalists~\cite{stray_making_ai_2019}. 

While there is generally limited empirical research on data journalists' workflows~\cite{chevalier_analysis_2018, stray_making_ai_2019}, among some extant process models related to producing data journalism, data preparation is an integral component under labels such as ``Clean''~\cite{bradshaw_inverted_2011, chevalier_analysis_2018} or simply ``Spreadsheets''~\cite{rogers_data_2013}. Skills for this stage are valued in the profile of a data journalist, yet secondary to traditional reporting skills~\cite{cardoso_practice_2022} and other aspects of data work, such as analysis and visualization skills~\cite{ojo_patterns_2018}. Rogers et al.~\cite{rogers_google_news_2017} \delete{find}\new{finds} that the prevailing view of data processing skills as a specialization is an organizational barrier that limit the use of data in newsrooms.

Many data journalism skills involve tools and techniques familiar in data science~\cite{stray_making_algorithms_2021, borges-rey_journalism_2021, showkat_where_2021}. Data journalists rely on general purpose tools, such as Excel and OpenRefine\footnote{Formerly known as Google Refine}~\cite{souza_interacting_2018} as well as specialized, open-source tools built by other journalists~\cite{pitts_open-source_2021}.

Our interviews provide substantially more detail about activities and issues when journalists prepare data, and our analysis carefully situates our findings with respect to the research literature. 

Showkat \& Baumer~\cite{showkat_where_2021} compares and contrasts practices in data-driven investigative journalism and data science. While our work shares a similar line of inquiry, it is distinct in two important ways. First, our work includes non-investigative reporting practices. Our participants describe their processes for both accountability reporting involving investigations as well as day-to-day reporting. Second, we recruit a broader participant pool of journalists, spanning multiple newsrooms. From this diverse perspective, we provide a broader characterization of the challenges data journalists face when preparing data. Our findings refute one claim of this work: we do not find that the use of unstructured documents vs structured data is a salient difference between data journalists and data scientists, respectively. Our participants often worked with structured data, including on investigative pieces.

Although no previous interview study addresses data journalists' workflows during data preparation, \delete{the TableScraps}\new{Table Scraps}\delete{work of Kasica et al.}~\cite{kasica_table_2021} is grounded upon a study of technical artifacts created by data journalists: their code notebooks and wrangling scripts. It answers questions related to \delete{\emph{what}}\new{what} journalists do when working with data through a taxonomy of \delete{\emph{actions}}\new{actions} and \delete{\emph{how}}\new{how} they do it through a taxonomy of \delete{\emph{processes}}\new{processes}. That study does not address the question of \delete{\emph{why}}\new{why} data journalists do what they do when preparing data, which our work seeks to answer by identifying issues inherent in raw data that affect its quality. Our taxonomy of actions does partially overlap with this previous model of activities, providing a complementary triangulation between knowledge gained from two sources: the artifacts journalists produce vs. their direct statements within interviews.

\begin{figure*}
  \includegraphics[width=\textwidth]{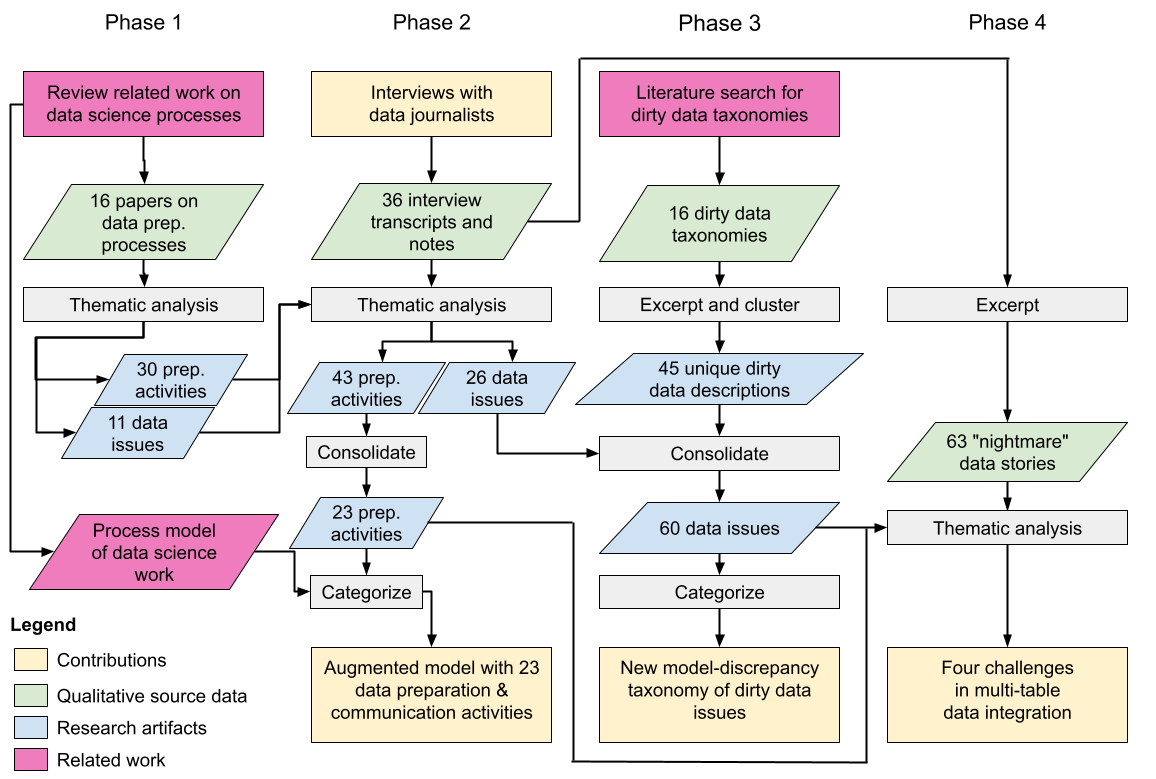}
  \caption{Process, products, and contributions: Our hybrid deductive-inductive thematic analysis~\cite{swain_hybrid_2018} began by analyzing \DSPaperCount~studies of data science workflows to generate \delete{\textnormal{a priori}}\new{\emph{a priori}} codes pertaining to data preparation (Phase 1). We then conducted an interview study with \N~data journalists on their preparation processes, generating \delete{\textnormal{a posteriori}}\new{\emph{a posteriori}} codes from those transcripts (Phase 2). The resulting artifacts yielded combined code sets of preparation \delete{\textbf{activities}}\new{activities} and data quality \delete{\textbf{issues}}\new{issues}. Our categorization of these activities extended a previous model of data preparation activities. We then analyzed \DQTaxonomyCount~taxonomies of dirty data issues (Phase 3), noting disparate coverage compared to our interview data. We produced a new model-discrepancy taxonomy for classifying dirty data issues to encompass them all. Finally, we reflected upon emergent patterns of data issues and preparation activities within the \delete{``nightmare stories''}\new{nightmare stories} section of our interviews to identify four challenges for \delete{multi-table} data integration (Phase 4).}
  \Description{A flowchart detailing the process of this study in four phases with individual steps depicted sequentally within each processes. Steps are colored if they related to our contributions, qualitative source data, research artifacts, and related work.}
  \label{fig:teaser}
\end{figure*}

\section{Methods} \label{sec:methods}

We conducted our study in four sequential phases:

\begin{enumerate}
    \item Analysis of data science workflow literature
    \item Analysis of novel interview data
    \item Analysis of dirty data issue taxonomy literature
    \item Further analysis of integration \delete{``nightmare stories''}\new{nightmare stories} from interviews
\end{enumerate}

The first two phases followed general guidelines for conducting hybrid thematic analysis, incorporating deductive approaches to create \delete{a priori}\new{\emph{a priori}} codes from previous work and inductive approaches to create  \delete{a posteriori}\new{\emph{a posteriori}}  codes~\cite{swain_hybrid_2018}. In the third phase, we analyzed an additional data corpus to contextualize our intermediate results\delete{, and in}\new{. In} the final phase\new{,} we further analyzed the \delete{``nightmare stories''}\new{nightmare stories} provided by \delete{interview}participants to identify four types of data integration challenges.  

\begin{table*}

\begin{tabular}{| l l | l | l l | }
\multicolumn{2}{|c|}{\textbf{\underline{Data science process papers}}} & & \multicolumn{2}{|c|}{\textbf{\underline{Dirty data taxonomies}}} \\
& & & & \\
\textbf{Study} & \textbf{Year} & & \textbf{Study} & \textbf{Year} \\

Kandel, Paepcke, Hellerstein, Heer~\cite{kandel_enterprise_2012}   & 2012 & & Chatterjee \& Segev~\cite{chatterjee_data_1991}       & 1991 \\
Kandogan, Balakrishnan, Haber, Pierce~\cite{kandogan_data_2014}    & 2014 & & Kim \& Seo~\cite{kim_classifying_1991}                & 1991 \\
Kim, Zimmermann, DeLine, Begel~\cite{kim_emerging_2016}            & 2016 & & Rahm \& Do~\cite{rahm_data_2000}                      & 2000 \\
Batch \& Elmqvist~\cite{batch_interactive_2018}                    & 2018 & & Dasu \& Johnson~\cite{dasu_exploratory_2003}          & 2003 \\
Kim, Zimmermann, DeLine, Begel~\cite{kim_data_2018}                & 2018 & & Kim et al.~\cite{kim_taxonomy_2003}                   & 2003 \\
Alspaugh et al.~\cite{alspaugh_futzing_2019}                       & 2019 & & M{\"u}ller \& Freytag~\cite{muller_problems_2003}     & 2003 \\
Battle \& Heer~\cite{battle_characterizing_2019}                   & 2019 & & Barateiro \& Galhardas~\cite{barateiro_survey_2005}   & 2005 \\
Kaggle~\cite{kaggle_state_2019}                                    & 2019 & & Oliveria et al~\cite{oliveira_formal_2005}            & 2005 \\
Mao et al.~\cite{mao_how_2019}                                     & 2019 & & Oliveria et al.~\cite{oliveira_taxonomy_2005}         & 2005 \\
Muller et al.~\cite{muller_how_2019}                               & 2019 & & Hellerstein~\cite{hellerstein_quantitative_2008}      & 2008 \\
Rule, Tabard, Hollan~\cite{rule_exploration_2018}                  & 2018 & & Li et al.~\cite{li_rule_2011}                         & 2011 \\
A.~Wang et al.~\cite{wang_human-ai_2019}                           & 2019 & & Gschwandtner et al.~\cite{gschwandtner_taxonomy_2012} & 2012 \\
D.~Wang, Mittal, Brooks, Oney~\cite{wang_how_2019}                 & 2019 & & Kandel et al.~\cite{kandel_profiler_2012}             & 2012 \\
Wongsuphasawat, Liu, Heer~\cite{wongsuphasawat_goals_2019}         & 2019 & & de Almeida et al.~\cite{de_almeida_taxonomy_2013}     & 2013 \\
Milani, Paulovich, Mannssour~\cite{milani_visualization_2020}      & 2020 & & Wickham~\cite{wickham_tidy_2014}                      & 2014 \\
Zhang, Muller, Wang~\cite{zhang_how_2020}                          & 2020 & & Roeder et al.~\cite{roeder_towards_2020}              & 2020 \\
\end{tabular}

\caption{\label{tab:related-work}\textbf{Related work}: (Left) In Phase 1, we analyze \DSPaperCount~data science process papers relevant to data preparation, a subset of those identified in a \delete{previous} systematic literature review~\cite{crisan_passing_2020}. (Right) We also analyze \DQTaxonomyCount~taxonomies of dirty data in order to better contextualize the data issues described by our participants.}
\end{table*}

\subsection{Phase 1: Data science workflow literature} \label{sec:methods-phase-1}

We constructed an initial codeset by analyzing accounts of data science workflows from previous interview, observation, and survey studies of data \new{scientists}. We began with the set of \DSPapersTotal~papers previously identified in a recent systematic literature review of data science workflows as being relevant to data preparation~\cite{crisan_passing_2020}. We excluded papers that do not directly derive their results from the lived experience of practicing data scientists. The remaining \DSPaperCount~papers that we analyzed are listed in Table~\ref{tab:related-work}; \SuppDSPapers~provides additional information on each of these papers, including study size, methods, and application domain.

These papers cover data scientists occupied in a diverse set of domains, described at different levels of abstraction. The most prominent domains were technology~\cite{kandogan_data_2014, kandel_enterprise_2012, milani_visualization_2020,muller_how_2019,kim_data_2018,kim_emerging_2016,wang_human-ai_2019,alspaugh_futzing_2019,wang_how_2019,wongsuphasawat_goals_2019}, including software engineering and social media; business~\cite{kandogan_data_2014, kandel_enterprise_2012,milani_visualization_2020,muller_how_2019,wang_how_2019,wang_how_2019, wongsuphasawat_goals_2019}, including finance; and healthcare~\cite{kandel_enterprise_2012, muller_how_2019, wang_human-ai_2019,alspaugh_futzing_2019,mao_how_2019,wongsuphasawat_goals_2019}. 

From each paper, the first author excerpted relevant sections on data preparation (resulting in \DSExcerpts~excerpts), then consolidated related excerpts into coherent groups through affinity diagramming~\cite{holtzblatt_contextual_2015}. Each group was given an \delete{\textnormal{a priori}}\new{\emph{a priori}} code. The resulting \DSCodes~codes were categorized into two higher-level \emph{families}~\cite{swain_hybrid_2018}: preparation \delete{\textbf{activities}}\new{activities} (\DSActivityCount) and data  \delete{\textbf{issues}}\new{issues} (\DSIssueCount).

\subsection{Phase 2: Novel interview data} \label{sec:methods-phase-2}

\delete{In the second phase, }We conducted \N~one-on-one, semi-structured interviews with data journalists from 31 different news organizations on their experience preparing data in the newsroom. We thematically analyzed these interview materials using the codeset generated deductively from related research in the previous phase, while generating new codes inductively from the interview data.

\subsubsection{Recruitment} \label{sec:recruitment}
To recruit \delete{interview}participants, we used purposive~\cite{robinson_purposive_2014} and snowball~\cite{frey_sage_2018}  sampling. We solicited interviews from a curated list of more than 100 contacts in our professional networks, considering the criteria of organization size (large, small), publication medium (print, broadcast, online), and business model (for-profit, non-profit, academic) in this purposive sampling to fill our participant pool with a representative cross-section of data journalists. We also used snowball sampling to request interviews from a few journalists (2/36) recommended by \delete{previous}participants. Because many data journalists do not have a formal job title connoting their expertise in data work, we used the inclusion criterion that participants should fit at least one of three personas:
\begin{itemize}
    \item \emph{Practitioner} (86\%): actively demonstrates data-oriented newswork through publishing articles, graphics, or applications at a media organization.
    \item \emph{Educator} (19\%): holds faculty or staff position at an institution of higher education and teaches classes on skills relevant to data journalism.
    \item \emph{Tool builder} (8\%): develops computational tools to assist in data-oriented newswork.
\end{itemize}
We did not use country as a criterion, but the final set of participants discussed experiences working at newsrooms based in Canada, India, the United Kingdom, and the United States. Full details on the \N~participants are available in \SuppParticipants.

\subsubsection{Procedure}
Prior to each interview, participants were asked to provide their informed consent and share artifacts related to specific data projects that were challenging with regard to preparing data. All participants complied, and this pre-interview background research primed the interviewer on subject material. The first author conducted each interview via video conference. See \SuppInterviewScript~for our interview script. All participants gave permission to record conversation audio, and the first author also took extensive notes during each interview. The average interview length was 49 minutes, and the \N~interviews yielded over 29 hours of recorded audio. 
The first author reviewed the recorded audio to revise the interview notes, transcribe salient portions as passages, and build familiarity with the data~\cite{swain_hybrid_2018}.

\subsubsection{Analysis} \label{sec:phase-2-analysis}
The first author applied the \DSCodes~\delete{\textnormal{a priori}}\new{\emph{a priori}} codes generated in the previous phase to appropriate passages and developed a total of \InterviewCodes~new \delete{\textnormal{a posteriori}}\new{\emph{a posteriori}} codes inductively from the interview data. After the final interview, the first author returned to earlier interviews to apply codes developed in subsequent interviews. For both kinds of codes, passages were selected that demonstrated qualitative richness~\cite{boyatzis_transforming_1998}. A total of \InterviewSnippets~passages were extracted and coded.

\subsubsection{Termination}
We concluded gathering data upon reaching theoretical saturation after 36 interviews, using the growth of our codebook's cardinality as a proxy for saturation. Notably, this number of participants conforms with sample size guidelines for qualitative studies with in-depth interviews~\cite{dworkin_sample_2012}. 

\subsubsection{Reflective synthesis} \label{sec:methods-phase-3}

We combined \delete{the \textnormal{a priori}}\new{\emph{a priori}} and \delete{\textnormal{a posteriori}}\new{\emph{a posteriori}} codes and searched for higher-level structure.

The \emph{activity} codeset contained \DSActivityCount~ \delete{a priori}\new{\emph{a priori}}  codes from data science workflow papers and \InterviewActivityCount~ \delete{a posteriori}\new{\emph{a posteriori}}  codes from the interviews, totalling \InitialActivityCount~activity codes (see \SuppActivities). Through reflective synthesis, we consolidated these into a final set of \ActivityCount~activity codes, renaming some for clarity. We realized that the activity codes could be used to extend the process model of data science work proposed by Crisan et al.~\cite{crisan_passing_2020} by adding an additional level of detail for the processes of data preparation and communication. We categorized all activity codes according to preparation \delete{sub-process}\new{subprocess}es (initiate, gather, create, profile, wrangle) and communication \delete{sub-process}\new{subprocess}es (disseminate, document). We present these results in Section~\ref{sec:activities}.

The \emph{issues} codeset contained \DSActivityCount~ \delete{a priori}\new{\emph{a priori}}  codes from previous workflow papers and \InterviewIssueCount~ \delete{a posteriori}\new{\emph{a posteriori}}  codes from the interviews, totalling \InitialIssueCount~issue codes (see \SuppIssues). Our first attempt at categorization through reflective synthesis did not lead to fruitful results. The high proportion of  \delete{a posteriori}\new{\emph{a posteriori}}  codes in this family (over 50\%) was one methodological indicator that the data sources in the first two phases contained highly divergent information. We thus chose to add another data source for the next analysis phase in hopes of bridging this gap. 


\subsection{Phase 3:  Dirty data taxonomy literature} \label{sec:methods-phase-4}

Many researchers studying data warehousing, data cleaning, and statistics have proposed taxonomies of dirty data. We reviewed this research literature with a snowball sampling approach. We started with a set of four such papers already familiar to us~\cite{kim_taxonomy_2003, kandel_profiler_2012, wickham_tidy_2014, chatterjee_data_1991}, then followed references and forward citations. We repeated this process until we discovered no further taxonomies of dirty data. Our final set of \DQTaxonomyCount~papers that contain dirty data taxonomies are listed in Table~\ref{tab:related-work}; \SuppDQTaxonomyCounts~enumerates the number of leaf nodes in the taxonomy trees, each corresponding to a dirty data issue we considered distinct. 

Our analysis collated \DQIssueCountRaw~concrete \delete{instances of dirty data}\new{\emph{instances of dirty data}}: the union of all leaf nodes in these taxonomy trees. We excluded \DQIssueCountExcluded~items that did not describe dirty data issues, were related to non-tabular forms of data, or whose descriptions we judged to be overly broad. We consolidated the remaining \DQIssueCount~issues by grouping together identical or essentially similar instances of dirty data into \ClusterCount~\emph{clusters}, listed in \SuppIssues. We then compare and synthesize these clusters of previously identified issues with our \InitialIssueCount~data issues from the previous phase, reconciling our self-generated labels to use existing terminology when applicable. This synthesis resulted in a set of \FinalIssueCount~issue codes, with \InterviewIssueCountUnique~unique to our interview analysis, \DQIssueCountUnique~unique to the previous work, and \IssueCountOverlap~overlapping. Our reflective synthesis of this material led to the new taxonomy for dirty data that we present in Section~\ref{sec:model-discrepancy-framework}. 

\subsection{Phase 4: Interview integration  \delete{``nightmares''}\new{nightmares}} \label{sec:methods-phase-5}

Finally, we conducted further analysis of the \delete{``nightmare''}\new{nightmare} stories told by the \N~participants, describing their difficulties combining data from multiple sources during data preparation. In this case, multi-table integration was the desired end, not a means to another end. All participants describe at least one such project, with \HorrorStories~in total across all interviews. There were \HorrorSnippets~coded passages from these stories, out of \InterviewSnippets~total passages extracted; see \SuppIntegrationChallenges~details. These passages had already been assigned activity and issue codes in Phase 2. Revisiting these passages, we found four emergent patterns occurring in \HorrorChallenges~out of the \HorrorStories~stories. We present and discuss these four data integration challenges in Section \ref{sec:integration}. 

\section{Preparation activities} \label{sec:activities}

\begin{figure*}
    \centering
    \includegraphics[width=\linewidth]{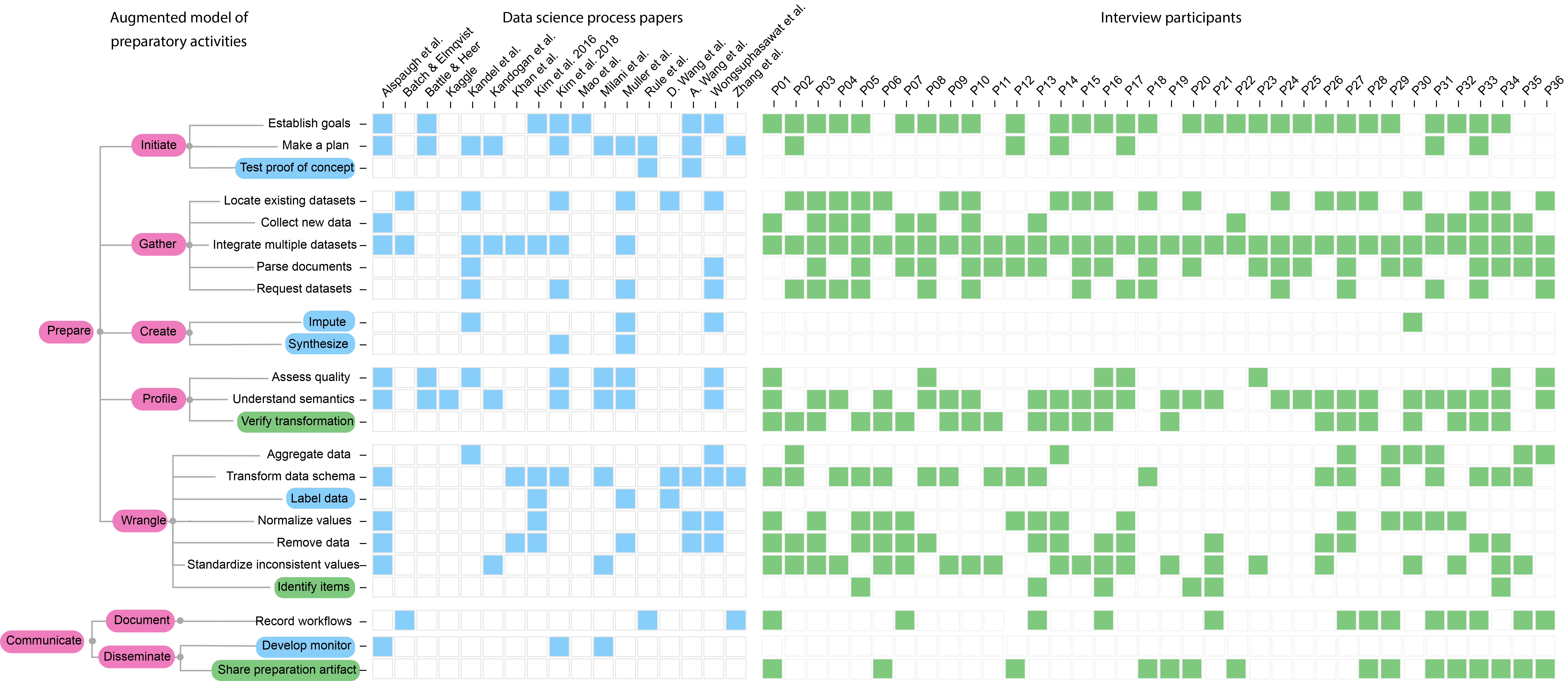}
    \caption{\textbf{Data preparation activities}: From our thematic analysis, we identify \ActivityCount~activities that data scientists and data journalists perform when preparing data; blue and green backgrounds highlight divergences. Legend: \testclr{RelatedPurple} processes from related work~\cite{crisan_passing_2020}, \testclr{DSBlue} prominent activities in data science, and \testclr{DJGreen} prominent activities in data journalism.}
    \Description{A heatmap showing the occurrence of both activities in our augmented model of preparatory activities among individual interview study participants (in blue), which were all data journalists, and research articles on data science workflows (in green).}
    \label{fig:data-prep-activities}
\end{figure*}

Our thematic analysis in Phases 1 and 2 results in a set of \ActivityCount~data preparation \delete{\emph{activities}}\new{activities}, shown in Figure~\ref{fig:data-prep-activities}. Our hybrid analysis approach allows us to distinguish between three cases, which we color code in figures and text: those performed by data scientists but not reported by our journalist participants (blue); activities that emerged directly from our journalist interviews that were not recorded in previous papers about data scientists (green); and activities performed by both groups (no highlight).


To increase the utility of our results, we categorize these activities within the process model of data science work proposed by Crisan et al.~\cite{crisan_passing_2020}. Their model posits four higher order processes (preparation, analysis, deployment, communication). Our activities all map to two of these: the preparation process and its five constituent \delete{sub-process}\new{subprocess}es (initiate, gather, create, profile, wrangle) and the communication process with its two \delete{sub-process}\new{subprocess}es (disseminate and document). By mapping activities to each of these, we extend their model to an additional level of detail.

\subsection{Initiate}

Data scientists often begin preparing a dataset by defining the needs of the project; outlining project objectives; and identifying requirements with colleagues, collaborators, and external stakeholders~\cite{crisan_passing_2020}. We call this process \emph{Initiate}, with the following three activities:

\begin{itemize}
    \item \emph{Establish goals}: define the overall objectives for a data project, including questions to answer, statistics to calculate, and final deliverables.

    \item \emph{Make a plan}: draft a proposal for a data project that specifies implementation details, monetary costs, and a rough timeline to achieve the established objectives.

    \item \activityDS{\emph{Test proof of concept}}: implement a small-scale test or pilot study before conducting a full-scale data project.

\end{itemize}

Our findings: We find that \activity{establishing goals} and \activity{making a plan} to achieve those goals can be challenging for data journalists when preparing an unfamiliar dataset. For example, the goal of generating new story ideas often requires a significant amount of data preparation, which may be prohibitively expensive. One participant explains:

\begin{quote}

Cleaning data takes so long, and here's the gamble: I don't know what the stories are in the data. But my track record indicates that there are stories in there...For a lot of media outlets that can't afford to free up people to do this kind of thing [data journalism], they're not necessarily going to take that gamble. If part of your pitch to your editor is 'I can spend weeks and weeks and weeks wrangling and cleaning this data, but I have no idea what the stories are.' Guess your odds of getting an editor to sign off on that? Virtually nil. That's the problem we face.

--- \McKie
\end{quote}

An unclear return on investment (ROI) is one barrier to the adoption of data journalism~\cite{rogers_google_news_2017}, and a few participants (3/36) lament time spent preparing data that did not yield publishable stories. In data science work, ROI can be clarified before investing significant time and resources by \activityDS{test proof of concept}. Notably, no participant describes conducting this activity. However, a few participants (2/36) describe identifying a ``minimum viable story'' in raw data with the expectation that further story ideas will appear during the preparation processes.

Throughout data preparation, journalists often discover limitations within their data that affect the goals initially established for the project. Many participants (15/36) report abandoning a data project due to issues such as cleanliness, complexity, and reliability; see \SuppAbandon~for details. One participant reports that a factor affecting their ability to achieve initial objectives is whether the data is an ``closed or open universe''. With a closed universe, they are sufficiently confident in the data's completeness to make absolute claims, but in an open universe they would always couch specific claims with disclaimers\new{,} such as ``at least''.


\subsection{Gather}
Data gathering includes the process of identifying existing data~\cite{crisan_passing_2020}. We expand this definition to include activities related to obtaining data. Both data journalists and data scientists perform these five gathering activities:

\begin{itemize}

    \item \activity{\emph{Locate existing data}}: find and identify data of interest either within their organization, publicly via the Internet, or from an external organization.
    
    \item \activity{\emph{Collect new data}}: record data from observed phenomena or processes in the world when existing data are not available.
    
    \item \activity{\emph{Integrate multiple data}}: combine multiple tables into one (including \emph{schema matching}~\cite{rahm_survey_2001}).
    
    \item \activity{\emph{Parse documents}}: create structured data by parsing data found in unstructured or semi-structured documents.
    
    \item \activity{\emph{Request data}}: request data from an organization, formally or informally.
    
\end{itemize}

Our findings: While both data scientists and data journalists \activity{request data} and \activity{parse documents}, every participant in our interview study reports issues that uniquely characterize these activities in data journalism.

Data scientists often work with data collected or maintained by clients or other divisions within a company, and may also make \activity{requests for data}. 
Many of our data journalist participants (14/36) \new{also} describe requesting data in a unique context not previously identified: formal data requests to government agencies through freedom of information (FOI) requests. It can take months or years for journalists to obtain data from FOI requests. These delays can lead journalists to abandon stories when they are no longer timely, \delete{and} thus less newsworthy. In response, many data journalists tend to gather data on phenomena that are newsworthy regardless of timeliness.

Often, data journalists receive data through FOI in PDF or physical documents, requiring them to further \activity{parse documents} in order to obtain usable data. Both data scientists and data journalists obtain these data by parsing unstructured or semi-structured documents, especially when scraping data from the web. However, parsing activities involving PDF documents, typically from FOI requests, is a unique context reported by our interview study participants. In this situation, journalists transform data populated in paper forms into tabular format or extract tables of data embedded in documents into a programatically accessible format.

Some participants (5/36) express the belief that FOI requested data has been deliberately returned in inaccessible forms that require extracting data, a sentiment that also been reported in other journalism studies~\cite{fink_data_2015}. One participant explains: 

\begin{quote}

Sometimes it's just what they're used to doing [supplying data in image-based PDFs], like they want to stamp it or they want to redact it. Sometimes I feel like they're just being ornery and don't want to be responsive to public information requests. I feel that way sometimes. I can't ever say that it's true, but I've definitely gotten data that way.

--- \McDonald
\end{quote}

All participants (36/36) integrate multiple tables, especially by supplementing one table with additional demographic data, such as local COVID-19 cases with demographic information from census data. Another common scenario is to detrend population-affected data by integrating data to calculate per capita rates, an established practice in precision journalism~\cite{meyer_precision_2002}. We discuss more challenging integration scenarios in Section~\ref{sec:integration}.

\subsection{Create}
When data cannot be collected or directly observed, data scientists may fabricate placeholder data~\cite{crisan_passing_2020}. From our analysis of the data preparation literature, we identify two activities within this subprocess, neither of which was prevalent with our journalist participants:

\begin{itemize}
    \item \activityDS{Impute}: replace missing data with values derived from other attributes.

    \item \activityDS{Synthesize}: fabricate data of hypothetical or approximate values that simulate data from observed phenomena.
\end{itemize}

Our findings: Almost no \activity{\new{data creation}\delete{create}} instances appear in our interview data. Every participant describes preparing data that represented observation of real-world processes or phenomena, but only one participant reports imputing missing values, in this case for six days out of an entire year. We attribute the extreme reluctance of journalists to impute or synthesize data to the professional norm to work only with material that might yield a publishable story and caution surrounding legal concerns if placeholder data accidentally appeared in print~\cite{berret_teaching_2016}.

Data journalists are also cautious about using data that contained estimates rather than observed values. One journalist discusses an instance investigating how the digital divide intersects with the COVID-19 pandemic as public education moved online, focusing on families in rural areas without access to high-speed internet. The journalist eventually abandoned the story because an official government dataset detailed hypothetical coverage rather than actual coverage, making it a poor measure of internet connectivity.

\subsection{Profile}
Profiling describes the \delete{sub-process}\new{subprocess} of assessing, understanding, and examining  data~\cite{crisan_passing_2020}. While checking for understanding is also a part of data exploration~\cite{alspaugh_futzing_2019}, we treat it as part of profiling due to the integral role it plays in other preparation processes, especially when removing data~\cite{grolemund_cognitive_2014}. We identify three profiling activities:

\begin{itemize}
    \item \activity{Assess quality}: ascertain the quality, identify issues, and any apparent limitations within a dataset.
    
    \item \activity{Understand semantics}: uncover or reveal the underlying meaning or context surrounding data.
    
    \item \activityDJ{Verify transformations}: ensure that recently applied data transformations did not have any unintended consequences. 
\end{itemize}

Our findings: With regard to profiling data, we note that data journalists exhibit a similar behavior to data scientists when assessing data, spending a significant amount of time understanding basic information about datasets, and using the same tools and techniques for other profiling activities to verify the effects of their transformations when wrangling and integrating data.

While visualization can be a powerful tool for assessing data, many data scientists assess their data numerically with summary statistics~\cite{milani_visualization_2020, kim_data_2018}. Many participants acknowledge that visualization could be useful in this activity\delete{,} but rely on numerical summary assessments of their data, such as counting the number of null values in an attribute.

Data scientists often devote significant time to understanding the nuances, underlying semantics, and subtle limitations of a dataset during data preparation. This activity is sometimes called ``becoming one with the data''~\cite{donoho_50_2017} or ``building intuition''~\cite{milani_visualization_2020}. Many participants (26/36) also stress the importance of developing a deep understanding of the dataset; they often spend significant time developing basic understanding because the data had inadequate documentation, if any. According to one participant:

\begin{quote}
Understanding, that's pretty big, especially when there's not enough documentation. You may request data but column names are 'ODCNLYTT' and you're like what is that? So there's a lot of incomplete documentation at all levels, but states, governments, have a way of either not providing or not properly documenting the data they collect in the first place.

--- \Kanik
\end{quote}

One new code to emerge from our journalist interviews involves \activityDJ{verifying} the effects of applied data \activityDJ{transformations} to confirm that no unintended side effects found their way into the transformed data. Our participants describe using profiling techniques to assess the quality of the transformed data. While they sometimes use visualization methods, they gravitate towards non-visualization methods, such as spot checks, summarizing attributes, and counting \delete{NULL}\new{null} or missing values. Journalists would also compare individual data items against previous versions of the same dataset to verify transformation effects.

\subsection{Wrangle}

Wrangling is defined elsewhere as the process of making data usable for analysis~\cite{kandel_research_2011}. However, as many other preparation \delete{sub-process}\new{subprocess}es are also aimed at this objective, we adopt a narrower definition of wrangling: modifying, refining, or otherwise altering a single table into an alternative form that is amenable to analysis. Many participants used synonyms for wrangling, such as \delete{\emph{munging}}\new{``munging''}, \delete{\emph{massaging}}\new{``massaging''}, and \delete{\emph{cleaning}}\new{``cleaning''}. We identify seven wrangling activities:

\begin{itemize}

    \item \emph{\activity{Aggregate data}}: decrease the size and granularity of a dataset by summarizing or grouping items in a table.
    
    \item \emph{\activity{Transform data schema}}: modify the underlying schema of the data.
    
    \item \emph{\activityDS{Label data items}}: annotate data items with semantically meaningful labels.
    
    \item \emph{\activity{Normalize values}}: adjust values measured on different scales to a common one.
    
    \item \emph{\activity{Remove data}}: decrease the size of the dataset by taking away, discarding, or filtering items or attributes from a table.
    
    \item \emph{\activity{Standardize inconsistent values}}: resolve inconsistency involving how the same entity is represented.
    
    \item \emph{\activityDJ{Identify items}}: distinguish unique items within a table or identify the same entities between multiple tables.

\end{itemize}

Our findings: While we find that data journalists mostly engage in the same wrangling activities as data scientists, we note that removing and normalizing data were especially prominent codes in our interview data in a context not addressed by related work on data science workflows. Additionally, \activityDJ{identify items} is a frequent and new code that emerged directly from our interview data. We speculate that this activity is not unique to data journalism, but is under-reported in data science workflows. Some data wrangling applications, such as Wrangler~\cite{kandel_wrangler_2011}, address this need through support for \delete{\emph{skolemization}}\new{skolemization}.

While the choice to \activity{remove data} often addresses noise, errors, and large datasets in data science, many journalists describe removing entire sections of extraneous data, items, and attributes that are not relevant to their inquiry during the initial steps of data preparation. ``I'll get a large dataset'', says \Cano, ``and 80\% of it is just stuff that I don't want''.

Some participants (6/36) describe creating unique keys \new{in a new attribute} to uniquely \activityDJ{identify items} or groups of items. Creating an attribute that identifies groups of items within a table is often a prerequisite activity for aggregating data within a single table. Journalists often craft this attribute as a soft key, with no guarantee of uniqueness~\cite{dasu_exploratory_2003}, by concatenating ostensibly unique attributes of names, addresses, birth dates, phone numbers. 

One participant describes encountering a table with what appeared to be duplicate data. The names and addresses matched. However, in reality they were father and son living in the same home, and the one differentiating attribute, birth date, was excluded from the dataset.

Data scientists often normalize data to satisfy model assumptions downstream in their workflow~\cite{wongsuphasawat_goals_2019, kim_emerging_2016}. Some participants (7/36) report normalizing \new{quantitative} data by calculating per capita rate, facilitating fair comparisons~\cite{meyer_precision_2002}. Some participants (7/36) describe normalizing qualitative data labels, mapping categorical attribute values into an ontology representing a different mental model. Participants rarely distinguish quantitative from qualitative data explicitly, but many journalists will create categorizations. For example, when preparing criminal justice data, one participant describes normalizing more than a dozen of a \new{court} judge's sentencing descriptions into three categories understandable by the general public, such as ``convicted'' or ``dismissed''. Arbitrary decisions around the definition of these labels can lead to differing conclusions in downstream analysis, as occurs in other phases of end-to-end data analysis~\cite{liu_paths_2020}.

Notably, no participant reports performing the activity \activityDS{label data}, or marking items as ground truth to train machine learning models, even though this activity is common in accounts of data science work. We believe preparing data for descriptive modeling, instead of predictive modeling itself, explains this difference. However, future work is needed to test this hypothesis.

\subsection{Document}
Documentation, or creating a record that describes performed work, is a communication process that intersects with data preparation and other high-level processes in data science~\cite{crisan_passing_2020}. While there may be other documentation activities across the entire data science process, \new{including archiving digital artifacts through documentation~\cite{heravi_preserving_data_journalism}}, we identify one documentation activity relevant specifically to data preparation:

\begin{itemize}
    \item \activity{Record workflow}: log the steps taken to prepare a dataset.
\end{itemize}

Our findings: Both data journalists and data scientists document aspects of data preparation by \activity{recording workflows}; however, data journalists contend with two distinct aspects of documenting the preparation process. First, in order to consolidate separate preparation processes performed across many different tools, some participants (5/36) created a \delete{workflow artifact called a ``data diary''}\new{\emph{data diary},} a separate document \new{containing data provenance information} typically composed \delete{in}\new{with} a word processor. While the data diary may include a list of steps made while preparing data, it may also include relevant preparation details beyond a simple data transformation log, such as data collection details, a contact phone number for questions about the data, and its limitations. Second, while data scientists often communicate their work to a variety of stakeholders~\cite{zhang_how_2020, wongsuphasawat_goals_2019}, data journalists focus on a unique stakeholder, the public. Thus, many data journalism articles post code, data, and methodological processes publicly~\cite{kasica_table_2021}.

\subsection{Disseminate}
Dissemination, or sharing insights into the data science process, is another cross-cutting data science communication process that intersects with data preparation~\cite{crisan_passing_2020}. We identify two activities where the two populations diverge:

\begin{itemize}
    \item \activityDS{Develop monitor}: create a means of checking the quality of a dataset as new items are ingested by the system.
    \item \activityDJ{Share preparation artifacts}: distribute byproducts of the data preparation process.
\end{itemize}

Our findings: Data scientists may \activityDS{develop} dashboards and other visualization artifacts to \activityDS{monitor} the data preparation process, often for dynamic datasets. However, only one of our journalist participants describes a single instance where they continuously maintained a dataset. Our participants \activityDJ{share artifacts}, executable snippets of code, intermediate data products for less technically adept colleagues, and reports on datasets after rounds of cleaning and vetting.

\section{Model-discrepancy taxonomy of dirty data issues} \label{sec:model-discrepancy-framework}

\begin{figure*}
    \includegraphics[width=.8\textwidth]{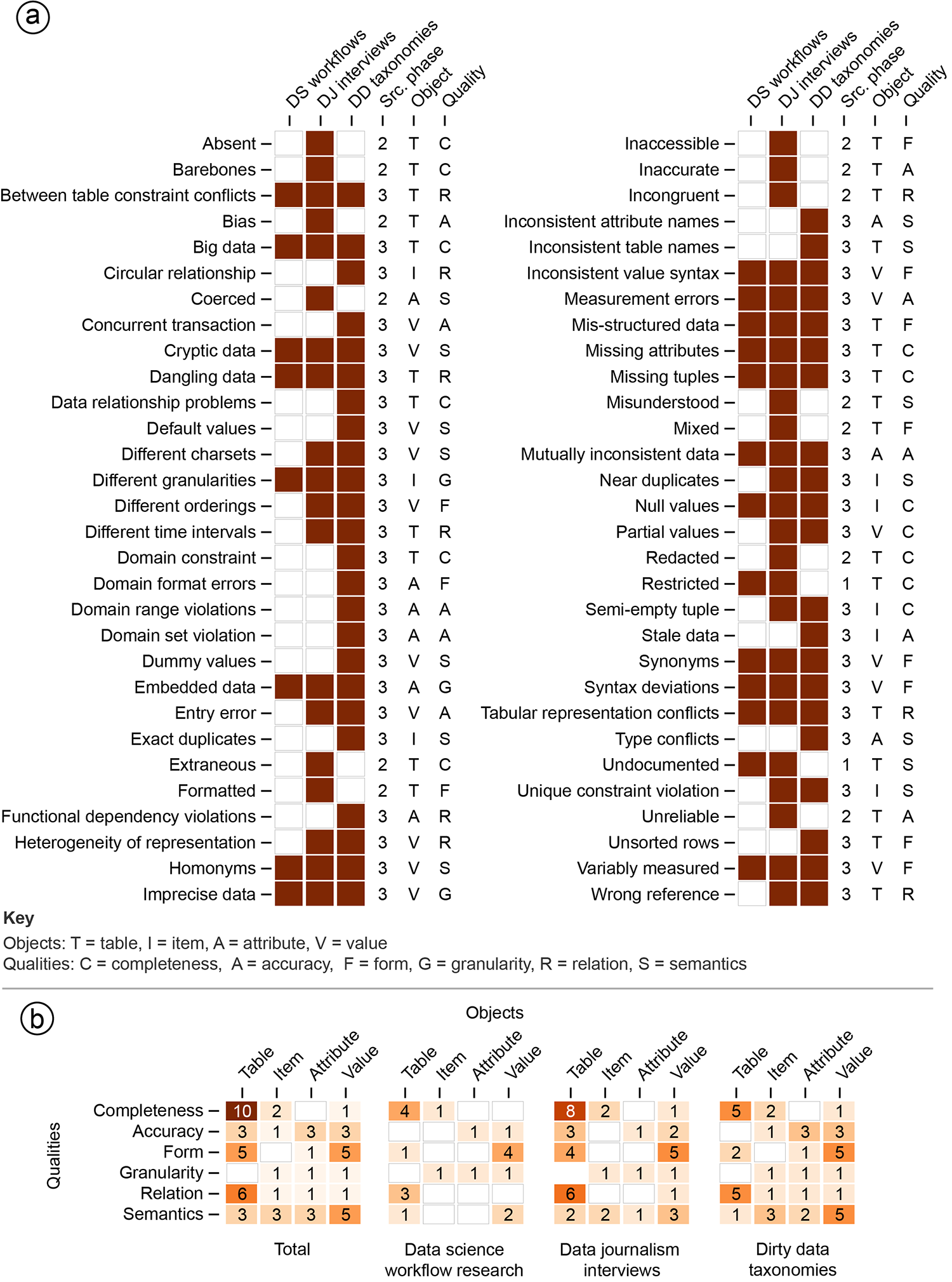}
    \caption{(a) \delete{60}\new{Sixty} data issues and which source of data they occur in (data science workflows, data journalism interviews, or dirty data taxonomies), the source phase they were identified in (1-3), and the object and quality the issue corresponds to within our model-discrepancy framework. See \SuppDataIssues~for a detailed explanation of each data issue. (b) The distribution of issues above in total and in each group of qualitative source data according to our new taxonomy for classifying dirty data.\delete{issues based on discrepancies between data workers' mental models of the data}}
    \Description{Sub-figure A show the occurrence of individual issue codes in which source of data they occur in: data science workflow research papers, data journalists interviews, and dirty data taxonomies. We also note which phase of our study these codes originate from and which object and quality they refer to in our model-discrepancy taxonomy of dirty data issues.}
    \label{fig:dirty-data-framework}
\end{figure*}

As we discuss in Section~\ref{sec:methods}, our initial attempt to categorize dirty data issues was unsatisfying. The Phase 1 material of workflow papers from the data science literature and the Phase 2 material of journalist interviews did not sufficiently overlap. We thus extended our analysis to include data quality issues discussed in the database literature, leading to a set of \FinalIssueCount~data quality issues that encompasses all three corpora of material, shown in Figure~\ref{fig:dirty-data-framework}. This large list requires some kind of hierarchical categorization to be useful, but previous data quality issue taxonomies lacked enough breadth to cover them all. Our novel taxonomy does, with a different lens than the others. 

Previous taxonomies of dirty data issues all characterize dirty data as falling short of some perfect kind of ideal data. In contrast, we view datasets as design artifacts made by data workers: people who collect, store, maintain, and prepare these datasets. Therefore, the data model represents the synthesis of mental models from the data workers involved. In this framing, dirty data constitutes an instance of a gulf where the existing data model does not match a data worker's mental model of the dataset. 

We propose a new taxonomy that classifies data quality issues as discrepancies between the user's model and the existing data model along two dimensions of a dataset: \delete{\emph{objects} and \emph{qualities}}\new{objects and qualities}. Both dimensions are orthogonal with regard to the specific data issues they categorize.

\subsection{Data objects}

The first axis of our taxonomy is built upon the four main concepts used by our participants to describe their data issues, using terminology following Munzner~\cite{munzner_visualization_2014}. It is considerably simpler than the more complex dirty data models from the database literature that handle multidimensional data. The four data objects are: 

\emph{Table}: a collection of items and attributes. Dirty data at the table levels affects multiple items and/or attributes. Tables are represented in rows and columns, but we use the term table to include other representations of tabular data, including a single relation in relational, JSON-structured, and XML data.

\emph{Item}: a collection of different attribute values that describe a specific observation or entities within a table.  Dirty data at the item level affects one or more attributes with regard to a single item. Items are uniquely identified through a combination of attribute values, a candidate key, or one unique attribute, a primary key. Equivalent terms: spreadsheet rows, tuple in a relation, or database records.

\emph{Attribute}: a specific, measurable property shared by items within a table. Dirty data at the attribute level affects multiple items along a single attribute. Unless explicitly specified in the design of the data schema through integrity constraint mechanisms, describable properties of attributes are often emergent qualities of the values associated with all items, including domain or range, semantic meaning, and associated data type. Equivalent terms: columns in spreadsheets and databases.

\emph{Value}: the amount or variety of a specific item with regard to a specific attribute in a table. Dirty data at the value level affects a single item-attribute pair. Values often carry many implicit assumptions that may not be expressed as attributes elsewhere in the table, such as units for quantitative measurement.

\subsection{Data qualities}
The second axis of our taxonomy contains six data qualities, \delete{namely} abstract characteristics of data objects. They are:

\emph{Completeness}: whether a data object has all the necessary and appropriate components. Instances of dirty data involving discrepancies in completeness can be characterized along a dual spectrum with opposing sides. Underlying missing data are discrepancies involving under-completeness. However, data with too much extraneous, irrelevant information also constitute discrepancies characterized by over-completeness. We concur with M{\"u}ller \& Freytag~\cite{muller_problems_2003} that removing incomplete data, instead of correcting the issue, artificially inflates the completeness of a dataset.

\emph{Accuracy}: the degree to which data objects are correct and precise with regard to the phenomena they represent in the world. While accuracy and precision are two separate measurements of observational error, we consider them together.

\emph{Form}: the arrangement, format, or configuration of data objects. Dirty data with discrepancies in form affect how data objects appear rather than what they mean or represent. Examples: pivot tables vs. tidy data~\cite{wickham_tidy_2014}; formatting attributes containing phone number, dates, or currencies; the order of attributes within a table.
    
\emph{Granularity}: a data object's scale or level of detail. As with completeness, the granularity of items and attributes may be above or below the expectation of a user's mental model. 
    
\emph{Relation}: the connection or relationship between multiple data objects of the same class. Examples: Multiple tables containing the same type of item~\cite{wickham_tidy_2014}; multiple attributes containing values with logical dependencies, such as ages and dates of birth.

\emph{Semantics}: the underlying meaning behind individual data objects. Undocumented or under-documented data can cause semantic discrepancies involving every data object. Attributes often carry high-level semantic types, such as people's names and social security numbers. These high-level types are often extensions of low-level types such as integers and character strings. Therefore, we consider conflicts involving primitive data types to be semantic discrepancies. Examples: multiple interpretations for the same value, such as homophones; conflicts between primitive data types, such as integer vs.~character string; and duplicate items.

Many of the data qualities we propose are not considered in previous models; the two that overlap with previous work are completeness and accuracy~\cite{muller_problems_2003, li_rule_2011, de_almeida_taxonomy_2013}. We describe many related, high-level classification schemes for dirty data in Section \ref{sec:related-dirty-data}.

\subsection{Results}

We note a substantial difference between the data issues that had received previous attention from database researchers and those uncovered by our interviews with data journalists: \DQIssueCountUnique~issues were unique to the database literature, \InterviewIssueCountUnique~issues were unique to data journalists' accounts, and only \IssueCountOverlap~overlapped. Figure~\ref{fig:dirty-data-framework} shows how this taxonomy covers these three groups of the \FinalIssueCount~dirty data issues, illustrating the need for a new taxonomy with adequate coverage of the full breadth of dirty data issues. 

We note that database researchers have a theory-focused perspective based on the concerns of people storing the data, whereas both data scientists and data journalists have a domain-oriented perspective focused on their data needs. We conjecture that the view of data issues found in our data journalist interviews may also pertain to other domain-oriented populations of people who use rather than store data. Our broad category of data workers encompasses all such consumers of data.

\section{Multi-Table Data Integration Challenges} \label{sec:integration}

In the final phase of our analysis, we re-analyze the \HorrorStories~ \delete{``nightmare''}\new{nightmare}  stories told by all \delete{interview}participants about their difficulties integrating multiple data sources. From these stories, we identify four recurring data integration challenges:

\begin{itemize}
    \item \delete{\emph{Regional}}\new{Regional}: tables with inconsistencies due to independent, spatially dispersed data sources.
    \item \delete{\emph{Diachronic}}\new{Diachronic}: tables on the same phenomena that evolve over time.
    \item \delete{\emph{Fragmented}}\new{Fragmented}: tables on a similar topic that contain different yet related items.
    \item \delete{\emph{Disparate}}\new{Disparate}: tables that are topically dissimilar and seemingly unrelated.
\end{itemize}

These challenges are not mutually exclusive; more than one challenge occurred in 11 of the \HorrorStories~nightmare projects. In an exceptional incident, one participant (\Bradshaw) reports using cloud computing services to process more than 10,000 individual tables for a story that involved a regional and diachronic dataset, combining a decade of monthly policing data from across the United Kingdom.

\subsection{Regional datasets}

Our participants report often working with open government data that federal regulatory or legislative bodies require be disclosed to the public, yet delegate the implementation of this mandate to constituent state, provincial, and municipal governments with little standardization guidance \delete{with regards to}\new{concerning} how this data is collected, organized, or disseminated. These constituent data collectors are often dispersed across disjoint geographic regions and institutional bodies. These conditions often produce \delete{\emph{regional}}\new{regional} datasets: multiple tables on the same phenomena from data collectors who are dispersed spatially and institutionally. Because many participants work on a level spanning the territory of multiple data collectors, our interviews reflect many issues with preparing regional data, including COVID-19 infection rates, political campaign expenditures, and crime statistics.

Due to the independence these data collectors exercise, a dataset may contain tables representing the same topic but structured differently in ways that impede the integration of these tables and make data preparation time consuming. One participant elaborates:

\begin{quote}
If I want to write a pan-Canadian story about a topic, it means I have to go to ten different provinces and ask them for data. No two of them will have the data in the same way ... There's different ways of recording the data and storing it. So to standardize this dataset into one single thing I can use takes a whole lot of time.

--- \Rocha
\end{quote}

However, local data journalists are also affected by regional data, sometimes to a greater degree than their national counterparts. One participant (\Jones) based in Missouri reports consolidating data from 90 municipalities to report on stories concerning a single county within the state.

The distributed nature of the dataset is often the most tractable issue with regional data, where many issues are perceived to stem from the independence of regional data stewards~\cite{crisan_passing_2020} in collecting, storing, and publishing data. One related issue is reconciling different classification ontologies for the same data items between multiple tables~\cite{kandel_enterprise_2012}. Our participants discuss reconciling incompatible ontologies on food hygiene ratings or types of business license. In our model-discrepancy taxonomy of issues, this data-related challenge represents a \discrepancy{value-relationality} discrepancy, and many participants describe the activity of \activity{standardizing} data in response. When standardizing dissimilar ontologies, a common strategy is to derive abstract categories that logically describe different categories. One participant describes reconciling different ontologies on use-of-force incidents by police departments across the United States:

\begin{quote}

Let's say a police department has 10 categories for use of force and another one has six...Deadly force is a less ambiguous category, but physical force that is non-lethal might be a broad category. Where one department has it broken down as like pushing and shoving and tasers and hitting with a baton, another department has it broken down as like push and shoving and everything else, you can then turn the three categories into one and be able to match them up.

--- \Arthur
\end{quote}

Participants report receiving regional data in many formats, including PDFs, spreadsheets, and flat text files, but also within email and sometimes values spoken over the phone. Other issues related to regional data are similar to those caused by data heterogeneity, conflicts in structure and representation arising from independently operated databases~\cite{kim_classifying_1991}. The data schema of one table may not conform to the data worker's mental model or the models of other tables. For example, data may be represented as one attribute or many. Some tables may be pivoted or cross-tabulated while others may be in tidy format~\cite{wickham_tidy_2014}. In the thorniest cases that participants describe, attributes may be intermittently present across tables, and the structure of multiple tables may not conform with the user's mental model.

The most insidious issues involve differences in data collection between regions that lead to \discrepancy{table-semantic} discrepancies regarding table items. For example, one participant (\Kanik) covering the opioid crisis found inconsistencies in the counts of fatal opioid overdoses due to different definitions of resident and cause of death.

\subsection{Diachronic datasets}

While difficulties surrounding the preparation of a regional dataset can be due to multiple, non-coordinating data sources on the same phenomena, a dataset published by the same source can still be difficult to prepare. Perennial news stories may involve analysis of civic data from a single source published at regular intervals, such as reports of government spending. However, data may evolve in subtle or dramatic ways in subsequent publications, leading to a \delete{\emph{diachronic dataset}}\new{diachronic dataset}: a set of tables on the same phenomena that structurally or semantically change over time. Some examples of diachronic datasets discussed by participants include: salaries of high-earning public sector employees, annual listings of donations to registered charities, and campaign contributions between election cycles. Some participants discuss preparing dynamic data, especially for dashboards and individual charts related to COVID-19 pandemic statistics. As the majority of participants discuss static data, we consider diachronic datasets a property of chiefly static data. However, many inherent issues may also extend to instances of preparing dynamic data.

Data issues associated with preparing diachronic datasets extend beyond discrepancies in \discrepancy{table-relationality}, stemming from data on the same phenomena separated into multiple tables~\cite{wickham_tidy_2014}. Changes due to the evolution of the dataset over time involve many other preparation \delete{sub-process}\new{subprocess}es, especially profiling~\cite{kandel_profiler_2012} and wrangling~\cite{kandel_wrangler_2011}.

Schema drift informally refers to changes in the data schema over time~\cite{noauthor_schema_2022}. This data issue represents a \discrepancy{table-form} discrepancy in our dirty data framework. Our participants report a common form of schema drift is the inclusion of additional attributes over time. As in data science, these attributes may be redundant~\cite{kandel_enterprise_2012}, but they may also represent new information. Moreover, participants describe addressing changing attribute names or meanings through \activity{transforming data}.

Another related issue involves the evolution of codes for a specific attribute. One entity may be referenced by two or more codes as the classification ontology evolves. Inflation is a common example involving quantitative data, and journalists derive index values to address this issue~\cite{meyer_precision_2002}. A more difficult issue involves the evolution of categorical value meanings, a form of \discrepancy{value-semantic} discrepancy. One participant (\Kanik) preparing economic data from the Bureau of Labor Statistics explains that while the occupation ``computer analysts'' is present in data from 1990, the meaning is not the same as in 2020.

Missing data is another common issue among participants, which we consider a \discrepancy{table-completeness} discrepancy, with regard to both attributes and items that represent continuous time periods. Some diachronic datasets may not be published at regular intervals, such as those released by hospitals. Other times, regularly published sources of data inexplicably dry up, according to a criminal justice reporter (\Pantazi) who analyzed prison population data. ``They're required by law to provide this data'', he says. ``But there's no punishment when they don't provide it''.

Participants report that anomalies within the data stem from undocumented methodological changes. One participant (\Weisz) says documented changes in the data collection methods are the exception rather than the rule. These methodological changes may result in anomalies may be detected when \activity{assessing} the data and require further \activity{understanding}.

Finally, participants describe a particularly difficult issue with shifting geographic boundaries for diachronic data representing a specific region. Cities grow. Smaller population areas amalgamate. Legislative districts are redrawn. These changes make fair comparison of the same area over time prohibitively complicated. A similar issue occurs when preparing \delete{\emph{fragmented datasets}}\new{fragmented datasets}, and methods used to address this issue may also apply to diachronic datasets.

\subsection{Fragmented datasets}

Both regional and diachronic datasets describe sets of tables with items that share the same meaning. However, another challenge many participants (17/36) describe involves preparing tables with items that are semantically distinct yet logically related: a \emph{fragmented dataset}. When \activity{requesting} data, many participants receive exported data that was previously organized into multiple related tables for efficient storage and retrieval. Examples of fragmented datasets include data on: rejected vote-by-mail applications and voter demographics; state hospitals and hospital procedures; and delinquent mine safety violations implicating multiple mines, operators, and owners. 

Preparing a fragmented dataset is like assembling a puzzle. The challenge involves \activity{understanding} how the pieces fit together. Many data preparation activities can involve combining a primary table with an auxiliary table containing an area's demographic or population data. What distinguishes preparing a fragmented dataset from standard data integration is that combined tables need not include all constituent components. Successfully prepared fragmented data may shed light on a particular aspect of the data, or it may reveal enough of the final picture to generate leads for traditional news reporting methods.

Fragmented datasets may have opaque codes from being originally stored in relational databases. Entities that represent categorical data may be represented as integers or other shortened codes in relational databases~\cite{chatterjee_data_1991}, and a related wrangling activity involves translating entity codes~\cite{kandel_enterprise_2012}. Journalists may approach resolving this issue as a form of \activity{standardization} or as an integration activity involving a lookup table, also known as a \new{\emph{crosswalk}}\delete{crosswalk}~\cite{woodley2008crosswalks}, a map that converts data to a new or different standard. This lookup table may have to be manually constructed by journalists from a data dictionary, textual descriptions for attributes accompanying published datasets~\cite{rashid2020semantic}. In some cases, the ``pieces'' may not align. Different tables may use different identifiers for the same entity, or the items in separate tables may represent overlapping, but not identical, geographic regions. 

Another area of difficulty is matching election results with demographic data, especially from national censuses. In many cases, demographic data must be \activity{aggregated} into larger areas equivalent to election precincts. However, some areas use idiosyncratic regions that census data cannot be aggregated into. One participant describes encountering this problem with Philadelphia's system of wards. 

\begin{quote}
This stuff isn't limited to just election data. The inability for different geographies to match up with each other is a well known problem that I think everyone who works with spatial data will encounter at some point in their lives, and we all have different ways of dealing with it.

--- \Zhang
\end{quote}

In this case, \Zhang~was able to address this issue by apportioning values by area, a technique used by an election blogger she consulted. Weighting overlapping regions by area or population may also be useful when integrating incongruent geo-political boundaries for the same areas, as seen when preparing diachronic datasets.

Reassembling related data accurately can be particularly challenging, even for veteran data journalists. Two participants coincidentally describe preparing data that combines the Debt by Age dataset and data on delinquent mine safety fines originally obtained by a FOI lawsuit filed against the US Mine Safety and Health Administration (MSHA). For the participant (\Jingnan) who prepared the original raw data from MSHA, under-documented and duplicate data impeded data preparation; \delete{while} an error in the received data added significant time to the preparation process. When one table was missing the ownership end date for some mines, it resulted in contradictory numbers that added months to the data preparation time. Later, the other participant (\Kanik), who works at a different news organization, used the cleaned Debt by Age dataset to report on safety fines from a specific mine owner. Despite being previously cleaned, this participant still encountered difficult aspects of preparing this data due to the complicated relationship between entities:

\begin{quote}

Delinquencies are applied to mines, but mines over time change ownership...if you're trying to find out who accrued the most violations in terms of ownership, you have to understand that you can't just pull from the violations and look at the owner's dataset or vice versa. They own mines that have violations that are unpaid that they're not responsible for, regulatorily.

--- \Kanik
\end{quote}

\subsection{Disparate datasets} \label{sec:disparate-datasets}

The three previous challenges describe datasets where individual items represent the same or similar topics. These challenges may involve spatial inconsistencies (regional), temporal variations (diachronic), or else inconsistencies that arise from idiosyncratic features of different source databases (fragmented). However, data journalism has a long history of gleaning insights by combining seemingly unrelated datasets~\cite{cohen_computational_2011, parasie_computing_2022}, and many participants (14/36) describe preparing data in order to integrate tables on dissimilar topics. These \emph{disparate datasets} are topically dissimilar, but contain reference to a common entity, such as attributes representing names, addresses, or phone numbers. These attributes can often serve as linkages between tables, and the intersection of these tables can reveal latent insights during an investigation, often implicating the subject in some form of wrongdoing.

The most common disparate dataset participants describe involves tables with items that semantically represented the same type of entity, such as people or companies, and specific items potentially referencing the same entity between tables. Hence, this process of combining multiple tables on common entities equates to entity resolution, reconciling multiple distinct references to the same real-world entity~\cite{kang_interactive_2008}. Interesting examples include:

\begin{itemize}
    \item Investigating healthcare workers dying of opioid overdoses by integrating tables of state health care provider licenses and death records.
    
    \item Identifying companies that laid off workers even though they were loaned funds from the federal Paycheck Protection Program designed to encourage small businesses to retain employees during the COVID-19 pandemic.
\end{itemize}

Many participants report two compounding issues when preparing disparate data that make it challenging to integrate datasets. First, there typically exist discrepancies in the identity of individual items. \activity{Identifying items} within a single table is a common preparation activity, which we identify in Section \ref{sec:activities}. When preparing disparate datasets, the difficulty of the activity is compounded by the additional requirement to craft keys that correctly reference the same entity between tables, serving as a makeshift foreign key. Second, inconsistent values can further impede data integration by complicating the process of creating keys. The same entity may go by multiple names, and different entities may use the same name. Hence, some journalists describe \activity{standardizing} data to reconcile inconsistencies, and a few describe circumventing standardization to some extent by relying on fuzzy match algorithms.

These issues can create uncertainty in the accuracy of match results. Journalists often deal with this uncertainty by re-evaluating the \activity{goals established} in the Initiate process. All journalists who report matching disparate datasets describe tuning their matching parameter in order to minimize or eliminate the rate of false positives in their combined data, which they acknowledge increases the percentage of false negatives in the results. Even a handful of correct matches can support multiple stories; however, publishing an incorrect match could end a career.

Disparate datasets may share another commonality that can be exploited to integrate two seemingly unrelated datasets: geography. While some instances of related data can be combined on equivalent geography, disparate datasets that are geographically related represent overlapping, but not equivalent, geographic regions. Often this type of disparate dataset involves census data, other data that do not use the same area definitions, or one area where the geographic boundaries change over time. A few participants (2/36) describe cases where this issue stopped them from pursuing a story, but one participant (\Zhang) describe resolving this issue though apportioning values by area. 

\section{Discussion} \label{sec:discussion}

We discuss \delete{the relationship between data journalists and data scientists,}the prominence of accountability journalism in our interviews, the role of tool usage in the capabilities of our \delete{interview}participants, and \delete{discuss}the implications of our work for the design of future tools.

\subsection{Accountability journalism}

Investigative journalists serve an essential role in democratic society, acting as a counterbalance to those who wield economic and political power by revealing corruption, dysfunction, abuse, and other forms of wrongdoing. 

We note that although data journalism may include other genres such as sports and entertainment reporting, participants in our study focused primarily on investigative journalism, also known as accountability or watchdog reporting. We conjecture that data preparation is most difficult for this type of journalism because it involves bringing transparency to unknown or deliberately concealed matters of concern. While sports organizations have an incentive to provide clean, readily usable statistics about games, teams, and players, corporations and governments often have the opposite incentive, leading to more laborious data preparation for journalistic investigations of business practices, health, labor, the environment, and other highly political subjects. 
 
Instances of wrongdoing by powerful figures and institutions are seldom readily apparent in the contents of a database or spreadsheet, especially when deliberate measures are in place to conceal this information. As a result, many works of investigative journalism require weeks, months, or even years of both traditional reporting and data-driven investigation, often straining the resources of news organizations whose budgets are already strained. 

\subsection{Tool-based archetypes and MacGyvering} \label{sec:arch-macgyver}

In Phase 1 and Phase 2 of our analysis, in addition to preparation \delete{\emph{activities}}\new{activities} and data quality \delete{\emph{issues}}\new{issues}, we also created a third family of codes for the usage of tools. We present these codes in \SuppTools. Our initial analysis did not yield the rich results of the other two code families, so we did not continue in search of formalisms. However, we did find one intriguing aspect of tool usage that both aligns and diverges from previous work, which we discuss here.  

Related work~\cite{kandel_enterprise_2012} proposes three archetypes of data worker based on the tools they use when performing data work: \emph{application users}, who use spreadsheets or other click-based applications (Excel); \emph{scripters}, who use software packages for data analysis (R or Matlab); and \emph{hackers} who are fluent in the same analysis packages as Scripters but also proficient in scripting languages (Python, Perl) and data processing languages (SQL).

Our participants align with this model of data worker expertise, especially with respect to how each archetype correlates with a set of tools used to prepare data. Many participants (13/36) were application users, employing only tools such as Excel, Google Sheets, or Microsoft Access. 

\delete{Many}Participants discuss working with colleagues who are proficient spreadsheet users but not data specialists, who would also fit into this archetype. Other participants fit the scripter archetype because they know basic Python (16/36). Finally, the most advanced users (7/36) were fluent in multiple programming languages and familiar with querying and creating databases, fitting with the hacker archetype.

However, in contrast to previous work~\cite{kandel_enterprise_2012}, we do not find that a data worker's preference for click-based vs. code-based tools necessarily restricted the preparatory activities they perform. We find that some application users implement a creative and improvisational approach to accomplishing preparation activities with the tools at their disposal. Following the term in widespread use among data journalists, we call this behavior \emph{MacGyvering}, after a 1985 American television series where the protagonist routinely escapes life-threatening scenarios through creative, even implausible, engineering feats using whatever objects happen to be nearby. One participant who fits the \delete{Application User}\new{application user} archetype describes the practice of MacGyvering in data journalism:

\begin{quote}
At some point, I might feel like the way I do this isn't sophisticated enough. I just don't know how to do this in a way that someone who knows how to program would. Are you just going to throw up your hands and give up? The point is not how beautiful your steps look. The point is, can you get there? Can you get there in a way that's accurate? If you have to MacGyver your way there with tape and spit, but it's accurate, then it's a success.
--- \SmithHopkins
\end{quote}

Application users MacGyver when re-appropriating existing data tools for unintended users. One participant uses tools for data removal that provide summary statistics to initially \activity{assess} and \activity{verify} data transformation as so called ``filter checks''. Some participants (5/36) without the experience to implement common data join operations supported in database applications or scripting languages still integrated data using copy-and-paste or chained calls to spreadsheet macro functions, such as \texttt{VLOOKUP}.

Similarly, some data journalists with enough technical expertise to satisfy the hacker archetype, and who predominantly manipulate data with scripting and database languages, will MacGyver when they incorporate click-based applications into their preparation process. A few participants \activity{standardize} data in OpenRefine due to the iterative control this application provides when performing this activity. One participant succinctly summarizes his reason for using multiple tools. ``I care about getting a story that somebody else doesn't have'', \Akin \space says. ``That's the job of journalists. I don't care what the tool is that lets me do it''.

\subsection{Implications for design}

Based on our results, we outline three recommendations for the development of data preparation tools that address the needs of data journalists.

\subsubsection{Support for verification activities} 

We find that \activity{verifying} the effects of recently applied data transformations is a profiling activity previously unidentified in the research literature. Participants describe using the same methods in other profiling activities to confirm that their mental model of a table matches the data model, via spot checks or visual exploration and assessment. All user archetypes engage in \activity{verifying} with a variety of tools.

From a design perspective, \activity{verifying} describes one way in which users attempt to understand the state of data throughout the preparation process. Hence, its presence reveals a gulf of evaluation~\cite{norman_design_2013} with regard to data states represented in preparation tools. With many tools used participants, this gulf is big. But designers could shrink this gulf by incorporating better feedback about the system state. Features that leverage data visualization can provide feedback at a scale that is easier to interpret.

\subsubsection{Support application users when integrating}
As mentioned in Section~\ref{sec:arch-macgyver}, we do not find the same limiting relationship between data worker archetypes and data preparation activities reported in related work~\cite{kandel_enterprise_2012}, especially concerning data integration. Despite not using the programming languages that implement join operations in relational algebra, application users who expressed \emph{MacGyvering} tendencies still integrate data through creative uses of available tools.

To better support data journalists, data preparation applications need to offer better support for combining data from multiple sources, especially those in our taxonomy of data integration scenarios (Section~\ref{sec:integration}). While many applications used by participants, such as Tableau Prep~\cite{tableau_prep}, do attempt to implement join operations, scalability is still an issue. The technical capability or interface design of these applications limits their usability when integrating data at the scale described by many participants.

We also find that proficient programmers will use applications for specific activities because they provide the same utility but greater usability than \delete{programming}\new{code}-based tools. Therefore, we believe users who currently prefer to integrate data using code-based tools may also consider applications for this activity if they offer the same utility but better usability.

\subsubsection{Support for preparation documentation}

Our study finds that both data scientists and data journalists create a record of data provenance when preparing data. While some preparation tools support provenance recording, version management, and workflow annotation, participants still perform this documentation activity in an external document, commonly called a data diary. This documentation artifact serves as a workaround in the absence of integrated data provenance tracking and reporting tools, and also reflects the extent to which participants used a very heterogeneous tool environment and could not rely on the internal capabilities of any single tool. We identify two limitations with the status quo that future tools should address.

First, preparation processes lack a cohesive medium to document workflow among the diverse set of preparation tools. Both data scientists and data journalists use a variety of tools when preparing data and often deploy their own idiosyncratic conventions for documenting data provenance. To the best of our knowledge, no system exists to ingest and unify provenance information from various applications, nor are there standards around the structure of data provenance information to promote interoperability between tools. Therefore, data workers must perform this consolidation manually in a word processor or text editor.

Second, data journalists report to a unique category of external stakeholder, the public, often publishing methodological posts documenting their data preparation in a so-called ``nerd box''~\cite{kasica_table_2021}. The step-by-step detail recorded in integrated documentation features used in data science are too granular for such public explanations, even for motivated citizens such as subject matter domain experts. Moreover, other stakeholders who are not directly working with the data such as supervising newsroom editors and collaborating journalists, also require a higher-level view of the data preparation processes. Journalists have a unique external audience in the form of motivated citizens, including subject matter domain experts, who may also be interested in high-level provenance information in the methodological posts that accompany many published instances of data journalism.

\section{Conclusion}


To understand how data preparation practices of data journalists compare to data scientists, we conduct an interview study with \N~data journalists and situate these results within research on data science workflows and dirty data. From these results, we propose a general taxonomy that considers data as a design artifact and dirty data as discrepancies between users' mental models, and we synthesize a process model of data preparation activities that data workers perform in pursuit of conforming data to one's mental model. We argue for the benefits of a more inclusive, pluralistic definition of data workers that includes both data scientists and data journalists. Although they perform many of the same preparation activities, we find important differences, including four challenges faced by journalists when combining \delete{multiple} tables during the preparation process: \new{regional, diachronic, fragmented, and disparate datasets}. Our findings can inform future work on the development of data preparation software. We encourage researchers to study and address the needs of all data workers, including data journalists.

\begin{acks}
We thank the UBC InfoVis group, especially Sam Fraser, for their useful feedback on paper drafts; our anonymous reviewers for their helpful comments; and the journalists who participated in our study for volunteering their time. This work is funded in part by \grantsponsor{1}{NSERC} \grantnum{1}{RGPIN-2014-06309} and the \grantsponsor{2}{Wallenberg AI, Autonomous Systems and Software Program (WASP)}.
\end{acks}

\bibliographystyle{ACM-Reference-Format}

\bibliography{interview-study} 

\end{document}